\documentclass[twocolumn,showpacs,preprintnumbers,amsmath,amssymb,APSl,prd,nofootinbib,superscriptaddress]{revtex4-1}

\usepackage{graphicx}
\usepackage{bm}
\usepackage{bbm}
\usepackage{amsmath}
\usepackage{amssymb}
\usepackage{amsthm}
\usepackage{color}
\usepackage{hyperref}
\usepackage{float}
\newcommand{\diff}{{\rm d}}

\newcommand{\mR}{\mathcal{R}}
\newcommand{\mRm}{\hat{\mathcal{R}}}

\newcommand{\LL}{\mathcal{L}}

\newcommand{\Mm}{\hat{M}}

\newcommand{\gm}{\hat{g}}

\newcommand{\hT}{h}

\newcommand{\Ht}{\tilde{H}}

\newcommand{\at}{\tilde{a}}
\newcommand{\nt}{\tilde{n}}
\newcommand{\Id}{\mathbbm 1}

\newcommand{\Omegah}{\hat{\Omega}}

\newcommand{\rhob}{\bar{\rho}}
\newcommand{\pb}{\bar{p}}
\newcommand{\mpl}{M_{\rm Pl}}
\newcommand{\mbi}{M_{\rm BI}}

\newcommand{\be}{\begin{equation}}
\newcommand{\ee}{\end{equation}}

\newcommand{\mS}{{\mathcal S}}

\newcommand{\bea}{\begin{eqnarray}}
\newcommand{\eea}{\end{eqnarray}}

\newcommand{\lsim}   {\mathrel{\mathop{\kern 0pt \rlap
  {\raise.2ex\hbox{$<$}}}
  \lower.9ex\hbox{\kern-.190em $\sim$}}}
\newcommand{\gsim}   {\mathrel{\mathop{\kern 0pt \rlap
  {\raise.2ex\hbox{$>$}}}
  \lower.9ex\hbox{\kern-.190em $\sim$}}}

\begin{document}

\title{On gravitational waves in  Born-Infeld inspired non-singular cosmologies}

\date{\today,~ $ $}

\author{Jose Beltr\'an Jim\'enez}
\email{jose.beltran@cpt.univ-mrs.fr}
\affiliation{Aix Marseille Univ, Universit\'e de Toulon, CNRS, CPT, Marse
ille, France}
\author{Lavinia Heisenberg}
\email{lavinia.heisenberg@eth-its.ethz.ch}
\affiliation{Institute for Theoretical Studies, ETH Zurich, Clausiusstrasse 47, 8092 Zurich, Switzerland}
\author{Gonzalo J. Olmo}
\email{gonzalo.olmo@uv.es}
\affiliation{Depto. de F\'{i}sica Te\'{o}rica \& IFIC, Universidad de Valencia - CSIC,\\
Calle Dr. Moliner 50, Burjassot 46100, Valencia, Spain}
\author{Diego Rubiera-Garcia}
\email{drgarcia@fc.ul.pt}
\affiliation{Instituto de Astrof\'{\i}sica e Ci\^{e}ncias do Espa\c{c}o, Faculdade de
Ci\^encias da Universidade de Lisboa, Edif\'{\i}cio C8, Campo Grande,
P-1749-016 Lisbon, Portugal.}


\date{\today}

\begin{abstract}
We study the evolution of gravitational waves for non-singular cosmological solutions within the framework of Born-Infeld inspired gravity theories, with special emphasis on the Eddington-inspired Born-Infeld theory. We review the existence of two types of non-singular cosmologies, namely bouncing and asymptotically Minkowski solutions, from a perspective that makes their features more apparent. We study in detail the propagation of gravitational waves near these non-singular solutions and carefully discuss the origin and severity of the instabilities and strong coupling problems that appear. We also investigate the role of the adiabatic sound speed of the matter sector in the regularisation of the gravitational waves evolution. We extend our analysis to more general Born-Infeld inspired theories where analogous solutions are found. As a general conclusion, we obtain that the bouncing solutions are generally more prone to instabilities, while the asymptotically Minkowski solutions can be rendered stable, making them appealing models for the early universe.
\end{abstract}

\maketitle

\section{Introduction}

After one hundred years of existence, General Relativity (GR) still stands out as the most succesful theory for the gravitational interactions \cite{Will:2014kxa}. However, this has not precluded an intense search for alternatives seeded by some challenges that GR currently faces (see \cite{Berti,MG} for some reviews). On large scales, the problems of dark matter (yet to be detected) and the cosmic acceleration (including the cosmological constant problem, yet to be solved) motivate infrared modifications of GR. In the ultraviolet sector, different modifications are contemplated as possible resolutions of cosmological and black hole singularities as well as the non-renormalisability of GR. These two problems are usually linked, as it is expected that quantum gravity effects will somehow regularise the singularities.

There is, however, another approach developed as a hope to avoid curvature singularities, which is rooted in the mechanism originally proposed by Born and Infeld as a way to regularise the electromagnetic divergences associated to point-like charged particles in classical electrodynamics \cite{BI}. The idea behind it is to replace the original Maxwell action by some square root structure so that electromagnetic fields cannot grow arbitrarily large due to the existence of an upper bound. The same idea was pursued by Deser and Gibbons to regularise curvature divergences in gravity \cite{Deser:1998rj}. However, a naive formulation in the metric formalism (where the affine connection is assumed to be \emph{a priori} given by the Christoffel symbols of the metric) introduces ghost-like degrees of freedom and, thus, it is not satisfactory. The situation improves if one uses instead a metric-affine (or Palatini) formulation of the theory \cite{Vollick:2003qp}, where the metric and affine connection are regarded as independent entities. In such a case, the ghost problem is naturally avoided and the theory does not introduce additional propagating degrees of freedom  (see \cite{PalOlmo} for a detailed account on the properties of gravitational theories in metric-affine formulation). Effectively, this formulation amounts to a modified coupling of standard matter fields to gravity. The most widely studied theory within this framework is the so-called Eddington-inspired Born-Infeld gravity (EiBI) and it is described by the action
\cite{Banados:2010ix}
\begin{eqnarray}
\mS_{\rm EiBI}&=&\mbi^2\mpl^2\int\diff^4x\Big[\sqrt{\det(g_{\mu\nu}+\mbi^{-2}\mR_{(\mu\nu)}(\Gamma))} \nonumber \\
&-&\lambda \sqrt{-g}\Big]\,,
\label{BIaction}
\end{eqnarray}
where $g$ is the determinant of the metric tensor $g_{\mu\nu}$, $\mR_{(\mu\nu)}(\Gamma)$ is the symmetric part of the Ricci tensor of an independent connection $\Gamma \equiv \Gamma^\alpha_{\mu\nu}$, $\mpl$ is the Planck mass, $\mbi$ is the scale at which modifications with respect to GR are expected to appear, and $\lambda$ is a constant related to the asymptotic nature of the corresponding solutions and encoded in the form of an effective cosmological constant that vanishes for $\lambda=1$ . This theory and its extensions have been extensively studied in the literature due to their many applications in cosmology, black hole physics and astrophysics \cite{BIapp} (see \cite{BeltranJimenez:2017doy} for a recent review on this class of theories).
A prominent feature of this theory is that it indeed succeeds in avoiding cosmological and black hole singularities in traditional scenarios such as FLRW cosmologies \cite{Banados:2010ix,oor14} and spherically symmetric configurations with electromagnetic \cite{Regular-Palatini} and scalar fields \cite{Scalar-Palatini}. Despite this preliminary success, it has been observed that the non-singular cosmologies suffer from instabilities, which casts doubts on the physical robustness of these solutions. In particular, instabilities associated with an unbounded growth of tensor perturbations in cosmological bouncing solutions were reported in \cite{EscamillaRivera:2012vz}. In subsequent works, the issue of these tensor instabilities in bouncing solutions was further considered \cite{Lagos:2013aua} and argued \cite{Avelino:2012ue} that, in some cases, such pathological behaviour could be avoided (see also \cite{Cho:2013pea}). 

In this work we revisit the behaviour of gravitational waves (GWs) in non-singular solutions from a general perspective. In section \ref{sec:BIgravity} we will review the non-singular solutions within the EiBI theory and summarise the governing equations of GWs, and analyse their behaviour in the presence of matter fields with constant equation of state parameter in section \ref{sec:tp}. We will see that one encounters a mild instability in the GWs propagation in the asymptotically Minkowski solutions of the EiBI theory whereas the bouncing solutions exhibit a more severe instability. Based on this result, we devote section \ref{roleofcs} to the role of the adiabatic sound speed of the matter source in the GWs propagation, since in Born-Infeld inspired gravity theories the background evolution is sensitive to the adiabatic sound speed of the matter fields. Fluids with more general equation of state parameter might yield new non-singular solutions but we will see that this comes at the price of a diverging adiabatic sound speed. In section \ref{moregenBI} we will perform a similar analysis for more general Born-Infeld inspired theories, for instance an extension in the form of a power law and elementary symmetric polynomials. Whereas the power law extensions share the same findings as in the EIBI theory, the symmetric polynomial extension will give rise to crucial differences concerning the asymptotically Minkowski solutions. We will summarise our findings in section \ref{conclusions} and mention a few interesting roads to be explored in future works. We will be working in the mostly plus convention and matrices will be denoted by a hat.

\begin{figure*}
\includegraphics[width=14cm]{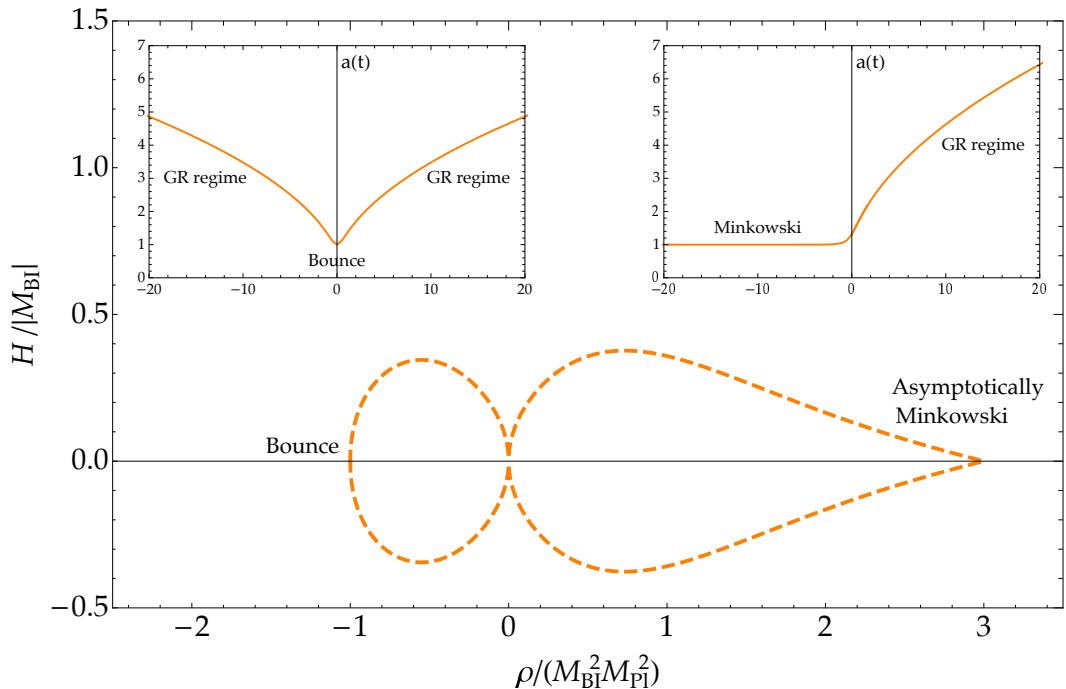}~~~
\caption{ In this figure we show the two non-singular solutions discussed in the main text with a radiation fluid as matter source. The modified Friedman equation given in Eq.(\ref{eq:H2}) is shown in the main figure where we can see that $H$ vanishes for $\rho=-\mbi^2\mpl^2$ and for $\rho=3\mbi^2\mpl^2$ (besides the usual Minkowski solution for $\rho=0$). In the inset plots we show the corresponding evolution for the scale factor near these non-singular solutions. In the case with $\mbi^2<0$ (left inset panel) we see the bouncing solution connecting a contracting phase with an expanding phase, both dominated by radiation. On the other hand, for $\mbi^2>0$ the scale factor approaches a constant value in the asymptotic past, while we have the usual solution for a radiation dominated universe with $a\propto \sqrt{t}$ at late times. }\label{Fig:EiBI}
\end{figure*}

\section{Non-singular cosmological solutions in EiBI gravity} \label{sec:BIgravity}

We will start by briefly reviewing the non-singular solutions within the EiBI theory, which will provide the basics for studying the problem of tensor perturbations instabilities at the bounce. By non-singular we refer to both bouncing and asymptotically Minkowski solutions, as we will discuss in more detail below. We will follow here the general procedure developed in \cite{BeltranJimenez:2017doy} to study cosmological solutions within this type of metric-affine theories. For later convenience, we will introduce the matrix
\be
\Mm\equiv\sqrt{\Id+\mbi^{-2}\gm^{-1}\mRm(\Gamma)} \ .
\label{def:M}
\ee
Since the theory is formulated in the Palatini formalism to avoid troubles with higher-order field equations and ghosts, the affine connection is an independent field whose explicit form has to be determined from the corresponding field equations. Under quite general circumstances (where, in particular, the connection does not enter into the matter sector), the connection can be algebraically solved as the Levi-Civita connection of an auxiliary metric $q_{\mu\nu}$, i.e., $\nabla_{\lambda}(\sqrt{-q}q^{\mu\nu})=0$. Moreover, such an auxiliary metric is related to the spacetime metric by means of a {\it deformation matrix}
\be \label{eq:relqg}
q_{\alpha\beta}=g_{\alpha\lambda}\Omega^\lambda{}_\beta  \,,
\ee
where $\Omega^\lambda{}_\beta$ is a function of the matter sector (and possibly the spacetime metric itself) determined from the metric field equations. The deformation matrix can be defined for a general theory \cite{BeltranJimenez:2017doy}, and for the EiBI theory it is simply given by $\Omegah=\hat{M}^2$.

Since we are interested in cosmological solutions we will assume the matter sector to be described by a homogeneous and isotropic perfect fluid whose energy momentum tensor is given by
\begin{equation} \label{eq:fluid}
T^\mu{}_\nu=\left(
\begin{array}{llll}
-\rho(t) &  &  & \\
 & p(t) &  & \\
 &  & p(t) & \\
 &  &  & p(t)
\end{array}\right)\,,
\end{equation}
where $\rho(t)$ and $p(t)$ are the energy density and pressure of the fluid, respectively. We will assume a minimally coupled matter sector so that the conservation of its energy-momentum tensor will be the same as in GR, i.e., we will have $\nabla^{(g)}_\mu T^{\mu\nu}=0$ for the covariant derivative corresponding to the Levi-Civita connection of the spacetime metric $g_{\mu\nu}$.  For this matter source, it is consistent to assume the matrix $\Mm$ to be also diagonal with $\Mm={\rm diag}(M_0,M_1,M_1,M_1)$. The metric field equations in such a case reduce  to
\bea
\frac{M_1^3}{M_0}=1+\rhob,\quad\,,
M_0 M_1=1-\pb
\eea
where we have introduced the re-scaled quantities $\rhob\equiv\rho/(\mbi^2\mpl^2)$ and $\pb\equiv p/(\mbi^2\mpl^2)$. Since $\mbi^2$ can be either positive or negative (describing two branches of solutions for the theory), the barred energy density can also take negative values  without running into physical pathologies. The above equations can be analytically solved as
\be
M_0=\left[\frac{(1-\pb)^3}{1+\rhob}\right]^{1/4},\quad M_1=\big[(1-\pb)(1+\rhob)\big]^{1/4} \,,
\ee
so that the components of $\Mm$ are algebraic functions of $\rho$ and $p$. These solutions already hint the existence of bounded solutions for the EiBI theory from the square root structure, as we will explicitly see shortly. The components of the deformation matrix $\Omegah$ (which will inherit the form of $\Mm$) can be easily computed and are given by
\bea
&&\Omega_0=M_0^2=\sqrt{\frac{(1-\pb)^3}{1+\rhob}} \label{eq:OmegaEiBI} \\
&&\Omega_1=M_1^2=\sqrt{(1-\pb)(1+\rhob)} \label{eq:OmegaEiBI1} \ .
\eea
In the cosmological scenario under consideration both metrics will have the FLRW form so that
\bea \label{eq:lineelg}
\diff s_g^2&=&-n^2(t)\diff t^2+a^2(t)\diff\vec{x}^2\\
\diff s_q^2&=&-\nt^2(t)\diff t^2+\at^2(t)\diff\vec{x}^2 \ ,
\eea
with $\{n,\nt\}$ and $\{a,\at\}$ the corresponding lapse and scale factor functions respectively. By using the form of the deformation matrix given in (\ref{eq:OmegaEiBI})
we find that the components of both metrics are related as follows: $\nt^2(t)=\Omega_0(\rho,p)n^2(t)$ and $\at^2(t)=\Omega_1(\rho,p)a^2(t)$.

As it is common in cosmology, we will now consider that the perfect fluid has a linear barotropic equation of state with constant parameter, i.e., $p=w\rho$ with $w$ constant. This is a very reasonable assumption for some standard cases like an ensemble of relativistic particles ($w=1/3$) or non-relativistic particles ($w\ll1$). We will turn to more general scenarios later and, in fact, this will have an important impact on the non-singular solutions of EiBI. Moreover, the conservation equation for the energy-momentum tensor in this scenario leads to the usual continuity equation
\be
\dot{\rho}+3H(\rho+p)=0 \ .
\label{eq:continuity}
\ee
The only dynamical matter variable is then the density $\rho$, which, according to (\ref{eq:continuity}), evolves as $\rho\propto a^{-3(1+w)}$ irrespectively of the gravitational Lagrangian. The general expression for the Hubble expansion rate, $H \equiv \dot{a}/a$, can be computed as\footnote{We will show the derivation of this equation for the general case below.}
\be
\frac{H^2}{\mbi^2 n^2}=\frac{3\Omega_0(M_1^2-1)-\Omega_1(M_0^2-1)}{6\Omega_1\left[1-\frac{3}{2}(\rho+p)\frac{\diff\log \Omega_1}{\diff\rho}\right]^2}\,,
\label{eq:H2}
\ee
where the right-hand-side is to be regarded as a function of $\rho$ determined by replacing $p=w\rho$ in the corresponding expressions of $M_0$, $M_1$, $\Omega_0$ and $\Omega_1$. At low energy densities, this expression reduces to the usual Friedman equation $\mpl^2H^2=n^2\rho/3$, while the differences appear at  energy densities of order $\mbi^2 \mpl^2$. This expression for the Hubble parameter has the usual zero for $\rhob=0$ giving the Minkowski solution, but it has two additional zeros at $\rhob=1/w$ for $\mbi^2>0$ and $\rhob=-1$ for $\mbi^2<0$ (see Fig. \ref{Fig:EiBI}). The first solution corresponds to an asymptotically Minkowski universe (\emph{loitering} solution), whereas the second case gives a \emph{bouncing} solution with a transition from a contracting phase to an expanding phase. Both of them correspond to scenarios where the scale factor reaches a minimum at some high energy, as depicted in Fig. \ref{Fig:EiBI}. At the asymptotically Minkowski point we have that $\pb=1$ so that $\Omega_0=\Omega_1=0$ and the auxiliary metric becomes singular. For the bounce we have that $\rhob=-1$ so that $\Omega_0$ diverges and, due to Eq.(\ref{eq:relqg}), the auxiliary metric diverges too. In both cases, the relation between the auxiliary and the spacetime metric is pathological. This does not need to be a problem as long as it has no physical effects, i.e., it might be enough that the spacetime metric (to which matter fields are sensitive)  is regular. However, although minimally coupled fields will only see the spacetime metric, tensor perturbations of the metric itself will be affected by this singular behaviour, as we explicitly show in the following, and this is at the heart of the difficulties for the construction of healthy non-singular solutions. In the remaining of this paper we will study this issue in detail and discuss the conditions to have stable non-singular solutions. We will conclude that one needs to go beyond this very simple scenario. But first, let us review the issue with the tensor perturbations.

\section{Evolution of gravitational waves in non-singular solutions in EiBI}\label{sec:tp}

Let us consider the tensor perturbations for the above non-singular solutions in EiBI gravity. We will introduce tensor perturbations for both metrics as follows: $\delta g_{ij}=a^2h_{ij}$ and $\delta q_{ij}=\at^2\tilde{h}_{ij}$. As shown in \cite{EscamillaRivera:2012vz}, the tensor perturbations for both the auxiliary and the spacetime metrics coincide, so we have $h_{ij}=\tilde{h}_{ij}$. This result was extended for a general class of Palatini theories  in \cite{Jimenez:2015caa}, where it was further showed that this is only true in the absence of any anisotropic stress. This is just a direct consequence of the deformation matrix satisfying $\Omega^i{}_j\propto\delta^i{}_j$, i.e., the space-space parts of the two background metrics are conformally related so that they have the same tensor perturbations. In that case, the corresponding evolution equation is given by
\be
\ddot{\hT}_{ij}+\left(3\Ht(t)-\frac{\dot{\nt}(t)}{\nt(t)}\right)\dot{\hT}_{ij}-\frac{\nt^2(t)}{\at^2(t)}\nabla^2 \hT_{ij}=0 \ ,
\label{htequation}
\ee
where we see that the tensor perturbations propagate on the auxiliary metric background.­ That this is the case can be easily seen by going to the {\it Einstein frame} of the EiBI theory. This frame for general metric-affine theories is carefully discussed in \cite{BeltranJimenez:2017doy} where we refer the reader for more details. In the present case, we can simply use the alternative bi-metric formulation of the EiBI theory given by the action
\bea
\mS_{\rm EiBI}&=&\frac12\mpl^2\int\diff^4x\sqrt{-q}\Big[q^{\mu\nu}\mR_{\mu\nu}(\Gamma)\nonumber\\
&&+\mbi^2\big(q^{\mu\nu}g_{\mu\nu}-2\big)\Big]+\mS_{\rm m}[\psi,g_{\mu\nu}] \label{eq:EiBIq}\,,
\eea
where $\mR_{\mu\nu}(\Gamma)$ is the (symmetrized) Ricci tensor constructed out of an arbitrary connection $\Gamma$. From Eq.(\ref{eq:EiBIq}) we see that the spacetime metric $g_{\mu\nu}$ only enters as an auxiliary field that can, at least in principle, be integrated out modifying the matter sector. The importance of this formulation of the EiBI action is that it is now apparent that the connection is determined as the Levi-Civita connection of $q_{\mu\nu}$ and its dynamics is governed by the usual Einstein-Hilbert term\footnote{More precisely, we obtain the Einstein-Hilbert term in the Palatini formalism which, as it is well known, is equivalent to its purely metric formulation.}. Thus, since the tensor perturbations on the FLRW background for both metrics coincide, the gravitational waves (GWs) will indeed see the background geometry determined by $q_{\mu\nu}$ rather than $g_{\mu\nu}$, as explicitly shown by Eq.(\ref{htequation}). Thus, even if the spacetime metric is perfectly regular, a singular behaviour in the auxiliary metric could be seen by gravitational waves. This is the origin of the tensor instability first discussed in \cite{EscamillaRivera:2012vz}.  To be more precise, we need the effective propagation speed and the friction term in (\ref{htequation}) to remain regular at the points where $H$ vanishes. To investigate this point in more detail, let us introduce the notation for these two quantities
\bea \label{eq:defcta}
c_T^2\equiv\frac{\Omega_0}{\Omega_1}\quad {\rm and}\quad
\alpha\equiv\sqrt{\frac{\Omega_1^{3}}{\Omega_0}} \ ,
\eea
so that the tensor perturbations equation can be written as
\be
\ddot{\hT}_{ij}+\left(\frac{\dot\alpha}{\alpha}+3H-\frac{\dot{n}}{n}\right)\dot{\hT}_{ij}-\frac{c_T^2n^2}{a^2}\nabla^2 \hT_{ij}=0 \ .
\label{htequation2}
\ee
It is interesting to notice that $c_T^2$ is positive if the deformation matrix is positive definite, i.e., we will not encounter any Laplacian instabilities whenever the two metrics share the same (Lorentzian) signature, which is the case for physically appealing solutions. Since the spacetime metric remains regular at those points where $H=0$, the potential divergences will be due to the behavior of $c_T^2$ and $\dot{\alpha}/\alpha$. These terms for the EiBI Lagrangian can be expressed in terms of the fluid variables by inserting (\ref{eq:OmegaEiBI}) and (\ref{eq:OmegaEiBI1}) into (\ref{eq:defcta}) to obtain
\bea
c_T^2&=&\left\vert\frac{1-\pb}{1+\rhob}\right\vert \label{eq:cT}\\
\partial_t\log\alpha&=&-3H(\rhob+\pb)\partial_{\rhob}\log\alpha=-3H\frac{\rhob+\pb}{1+\rhob}\,, \label{eq:alpha}
\eea
where in the last equation we have considered $\alpha$ as a function of $\rho$ only, with $p=w\rho$, and we have used the continuity equation (\ref{eq:continuity}).

\subsection{Gravitational waves instabilities}

Now it is straightforward to analyse the behaviour of the tensor perturbations at those points:
\begin{itemize}
\item Bouncing solution. For this solution we have $\rhob=-1$ so we see that $c_T^2$ diverges, signaling a singular behaviour of the tensor perturbations. On the other hand, near the bounce we have $H^2\propto (1+\rhob)$ so that $\dot{\alpha}/{\alpha}\propto (1+\rhob)^{-1/2}$ and thus the friction term also diverges at the bounce.
\item Asymptotically Minkowski. In this solution we have $\pb=1$ so that we obtain $c_T^2=0$ and $\dot{\alpha}/\alpha=0$, hence Eq.(\ref{htequation}) is approximately $\ddot{h}_{ij}\simeq0$ and, thus, we have that tensor perturbations grow as $t$ for $t\rightarrow-\infty$, i.e., in the asymptotically Minkowski region.
\end{itemize}
We have re-obtained the results of \cite{EscamillaRivera:2012vz} in a slightly different way. However, here we did not need to solve the time evolution of the metrics but instead to put forward the direct relation between the singular behaviour of the auxiliary metric and the presence of tensor instabilities. We should emphasize that the instability in the bouncing solution looks more severe than the one obtained for the asymptotically Minkowski case, where the instability amounts to the mildly growing mode linear in time. In particular, if we start with an initial state at some large $\vert t\vert$ with a finite amplitude for the tensor perturbation, the above result would suggest that the universe would actually isotropize with the expansion. However, the bouncing solution is much more dramatic since the divergence occurs exactly at the bounce and, moreover, it remains even for the long wavelength modes. This seems to signal the presence of a shear instability at the non-linear level. To see that this is actually the case, we will now turn to the non-linear case to understand the rise of a shear instability at the bounce, while the asymptotically Minkowski solution is free from shear instabilities.

\begin{table}
\begin{center}
\begin{tabular}{|c||c|c|}
\hline
&Bouncing ($\rhob=-1$)& Asymptotically Minkowski ($\pb=1$)\\
\hline\hline
$\alpha$&$\vert1+\rhob\vert\rightarrow0$&$\alpha\to \vert1+1/w\vert$\\
\hline
$\dot{\alpha}/{\alpha}$&$\vert1+\rhob\vert^{-1/2}\rightarrow\infty$&$\vert1+\pb\vert\rightarrow0$\\
\hline
$c_T^2$&$\vert1+\rhob\vert^{-1}\rightarrow\infty$&$\vert1+\pb\vert\rightarrow0$\\
\hline
\end{tabular}
\end{center}
\caption{In this table we summarise the behaviour of the relevant parameters for the evolution of the GWs in EiBI theory near the non-singular solutions discussed in the main text.}
\label{Table}
\end{table}%

\subsection{Non-linear instability}

We have seen that both of the above non-singular solutions are unstable because the singular behavior of the auxiliary metric $q_{\mu\nu}$ is transferred into a divergent growth of tensor perturbations. A natural question then is the nature of this divergence at the non-linear level. For that purpose, we will consider a homogeneous but anisotropic metric of the Bianchi I form
\be
\diff s^2=-n^2(t)\diff t^2+a_1^2(t)\diff x^2+a_2^2(t)\diff y^2+a_3^2(t)\diff z^2.
\ee
This problem was already treated in \cite{Olmo:2012yv} for a class of Palatini theories. In that work, it was shown that the shear is given by
\be
\sigma^2=\left(\frac{C}{3\Omega_0V}\right)^2\,,
\ee
where $C$ is a combination of integration constants and $V\equiv (a_1a_2a_3)^{1/3}$ is the volume of the congruence. We then see that the shear becomes singular for the bouncing solution since $\Omega_0$ vanishes at the bounce. However, the asymptotically Minkowski solution leads to $\Omega_0=1+1/w$ so that there is no shear divergence (at least for a radiation fluid, which is the case we are interested in throughout this work). Thus, we find again that the instability for the bouncing solution is more severe than the one for the asymptotically Minkowski case.

\subsection{Gravitational waves in the asymptotically Minkowski phase.}

The results of the precedent sections suggest that the asymptotically Minkowski solution is not as pathological as the bounce and, in particular, GWs present a smoother behaviour. For this reason, it is interesting to consider more carefully the behaviour of the GWs near the asymptotically Minkowski phase. As it is apparent from Fig. \ref{Fig:EiBI}, the universe remains in the Minkowski phase from $t=-\infty$ until the GR regime is reached and we enter the usual expanding universe dominated by radiation. Thus, in the Minkowski phase $H\simeq0$ and the GWs evolve according to
\be
\ddot{\hT}_{ij}+\frac{\dot\alpha}{\alpha}\dot{\hT}_{ij}-c_T^2\nabla^2 \hT_{ij}=0 \,,
\label{htequationH0}
\ee
where we have normalised the scale factor to its value at $t=-\infty$, so $a\simeq 1$, and we have chosen cosmic time, so that $n=1$. This simplified equation, which is valid throughout the entire Minkowski phase at high densities, is very illuminating to understand the behaviour of the GWs. We see that the friction term $\dot{\alpha}/\alpha$ is simply telling us that the GWs will see a cosmological expansion different from the one to which matter fields are sensitive to. Thus, while matter sees an effective Minkowski space, GWs evolve as if they were in an effective FLRW universe with a Hubble factor proportional to $\dot{\alpha}/\alpha$ and with a propagation speed given by $c_T^2$. This immediately allows us to understand that the actual pathology for the GWs around these non-singular solutions is the vanishing of $c_T^2$. Having $\dot{\alpha}/\alpha=0$ simply signals that GWs will also see an effective Minkowski background space. The vanishing of $c_T^2$ is however more problematic since this might point towards a strong coupling problem in that regime. An easy way to understand this is that, with $c_T^2=0$, exciting spatial gradients of GWs costs no energy at all. In order to be more precise, we can write down the quadratic action for the GWs. This is a trivial task if we work in the Einstein frame discussed above where it is apparent that GWs will have the usual Einstein-Hilbert term so that
\be
\mS^{(2)}_h=\frac{\mpl^2}{8}\int\diff^3x\diff t \frac{\at^3}{\nt}\left(\dot{h}_{ij}^2-\frac{\nt^2}{\at^2}\vert\nabla h_{ij}\vert^2\right).
\ee
Obviously, this action leads to the GWs equations given in (\ref{htequation}). If we evaluate it near a non-singular solution with $a\simeq 1$ we have the expected result
\be
\mS^{(2)}_h=\frac{\mpl^2}{8}\int\diff^3x\diff t \alpha\left(\dot{h}_{ij}^2-c_T^2\vert\nabla h_{ij}\vert^2\right),
\ee
where it is apparent that $\alpha$ plays the role of an effective scale factor\footnote{ To be more precise, $\alpha=a_{\rm eff}^3$ in the corresponding cosmic time or $\alpha=a_{\rm eff}^2$ in conformal time.}. We thus see that $\alpha$ simply amounts to a change in the normalisation of the GWs and, moreover, it is $\alpha$ that has a precise physical meaning. Let us notice that, by the own definition of $\alpha$, this is a positive quantity and, thus, no ghost instabilities will ever arise, as one would have expected. Now, if we canonically normalise the GWs as $h_{ij}=4h^{c}_{ij}/(\mpl\sqrt{\alpha})$ then the interactions of the GWs will be suppressed by a scale $\sim \sqrt{\alpha}\mpl$ so that if $\alpha$ vanishes this scale is arbitrarily small and the interactions will become very relevant, breaking down the perturbative expansion. This is precisely what happens for the bouncing solutions. However, the asymptotically Minkowski solution gives a finite value for $\alpha$ so that this strong coupling problem is absent.

Concerning $c_T^2$, we can choose a time coordinate $\tau$ such that $\diff \tau=c_T\diff t$ so that the action can be expressed as
\be
\mS^{(2)}_h=\frac{\mpl^2}{8}\int\diff^3x\diff \tau\, \alpha c_T\Big[(\partial_\tau h_{ij})^2-\vert\nabla h_{ij}\vert^2\Big].
\ee
This form of the action shows an interesting face of the instability. Let us notice by passing that the product $\alpha c_T$ is finite for the bouncing solutions (see Table \ref{Table}) so, neglecting for a moment the very unpleasant presence of a divergent propagation speed prone to causing causality problems, one might be tempted to think that GWs behave smoothly in this time coordinate with no strong coupling problems. However, this time re-scaling will also affect the interactions of the GWs and the matter sector and, therefore, the strong coupling problem has not disappeared but translated into those sectors. In the asymptotically Minkowski solution however, $\alpha$ remains finite while $c_T^2$ vanishes and we can then clearly see the strong coupling problem arising from the vanishing of $c_T^2$, while the vanishing of the friction term in the GWs equation simply signals that they also see an effective Minkowski space, but that does not introduce any pathology.

The type of behaviour found for the GWs with vanishing propagation speed is also encountered in other models of modified gravity like the ghost condensate \cite{ArkaniHamed:2003uy}. The resolution to this difficulty in that case is that the gradient energy will then come from higher derivative interactions that would stabilise the system. Within an effective field theory perspective, we could also expect the presence of higher derivative interactions in the Lagrangian of the form\footnote{This type of interactions could be expected to arise from terms like e.g. $R_{\mu\nu}R^{\mu\nu}$ in the Einstein frame. We should bear in mind however that the original theory is formulated in a metric-affine framework and the appearance of such terms should be analysed with more care as well as the consistency of the full theory as an effective field theory. For instance, it is expected that terms involving derivatives of the curvature, i.e. interactions involving $\partial_\lambda \mR^\alpha{}_{\beta\mu\nu}$, will be required to obtain the desired higher derivative terms in the quadratic action.} $(\nabla^2 h_{ij})^2$ that would give the appropriate dispersion relation with an analogous stabilisation mechanism. By including those terms and taking $c_T\rightarrow0$ the quadratic action for GWs would read
\be
\mS^{(2)}_h=\frac{\mpl^2}{8}\int\diff^3x\diff t \alpha\left[\dot{h}_{ij}^2-A (\nabla^2 h_{ij})^2\right],
\ee
with $A$ some (possibly time-dependent) parameter. This quadratic action leads to a dispersion relation of the form $\omega^2\propto k^4$ instead of the usual quadratic relation. The vanishing of the sound speed for the perturbations around a non-trivial background also occurs for the transverse phonons in the effective field theory of fluids \cite{Endlich:2010hf}. In that case, even higher derivative interactions will not resolve the problem because this feature is actually imposed by a symmetry. In the case of the EiBI there is no symmetry imposing $c_T^2=0$ at the asymptotically Minkowski phase and, thus, it is closer to the case of the ghost condensate model.

From this discussion it is now clear that the actual pathology for the behaviour of GWs in asymptotically Minkowski solutions is the vanishing of $c_T^2$, but the problem could be resolved by including higher order derivative terms. However, unlike the problems arising in the bouncing solutions, the vanishing of the propagation speed for the asymptotically Minkowski phase can actually be avoided within other Born-Infeld inspired gravity models, as we will explicitly see for a specific theory in Section \ref{Sec:polynomial}.

\section{The role of $c_s^2$}\label{roleofcs}

In the previous section we have discussed the presence of tensor instabilities in bouncing and asymptotically Minkowski solutions for the EiBI model. We have seen that, while the latter gives rise a mild instability in the asymptotic past, the former exhibits a more severe instability caused by a divergence of the propagation speed and the vanishing of the GWs normalisation $\alpha$. Those results have been obtained by assuming a barotropic fluid with constant equation of state parameter. However, in Born-Infeld inspired theories (and in metric-affine theories in general) the cosmological evolution also depends on the {\it adiabatic sound speed} of the matter source. This is a crucial difference with respect to most modified gravity theories that we want to explore here to see if it can help alleviating the tensor instabilities. In fact, this was already explored in \cite{Avelino:2012ue} for a fluid with time-dependent equation of state where the authors argued that such a time dependence could resolve the presence of tensor instabilities in bouncing solutions. Here we will consider a more general scenario in the presence of a fluid with a general sound speed and show the mechanism by which the GWs could be rendered stable in bouncing solutions. As we will see, this generically relies on a divergent sound speed at the bounce. For the sake of clarity, it will be convenient to briefly go through the crucial steps for the derivation of the modified Friedman equation  (see \cite{BeltranJimenez:2017doy} for a more detailed derivation).

As usual, we start by computing the $00$ component of the Einstein tensor of the auxiliary metric from its definition
\be
G_{00}(q)=3\left(\frac{\diff \log\at}{\diff t}\right)^2=3\left(H+\frac12\frac{\diff \log\Omega_1}{\diff t}\right)^2 \ ,
\ee
where we have used that $\at^2=\Omega_1 a^2$. Now we need to notice that $\Omega_1$ will still be the function of $\rho$ and $p$ given in (\ref{eq:OmegaEiBI}), as dictated by the metric field equations. Then, we can expand its time derivative as\footnote{Let us notice that the expressions used throughout this derivation are insensitive to a re-scaling of the density and the pressure, so that all formulae are the same for both the barred and unbarred variables.}
\be
\frac{\diff \log\Omega_1(\rhob,\pb)}{\diff t}=-3H(\rhob+\pb)\Big[\partial_{\rhob} \log\Omega_1+c_s^2 \partial_{\pb} \log\Omega_1\Big] \ ,
\ee
where we have defined the {\it adiabatic sound speed}
\be
c_s^2\equiv\frac{\dot{\pb}}{\dot{\rhob}} \ ,
\ee
and we have made use of the continuity equation (\ref{eq:continuity}). We can now use this result to re-write $G_{00}(q)$ as
\be
G_{00}(q)=3H^2\left[1-\frac32(\rhob+\pb)\Big(\partial_{\rhob} \log\Omega_1+c_s^2 \partial_{\pb} \log\Omega_1\Big)\right]^2 \ .
\ee
On the other hand, we can alternatively express the Einstein tensor
\be
G_{\mu\nu}(q)=R_{\mu\nu}(q)-\frac12\big(q^{\alpha\beta} R_{\alpha\beta}\big) q_{\mu\nu} \ ,
\ee
in terms of $\hat{M}$ and the deformation matrix $\Omegah$ from their definitions in (\ref{def:M}) and (\ref{eq:relqg}). Then, we have
\be
G_{00}=\frac12\mbi^2g_{00}\left[M_ 0^2-1-3\frac{\Omega_0}{\Omega_1}\big(M_1^2-1\big)\right] \ .
\ee
Equipped with this result we can finally obtain the modified Friedman equation that generalises (\ref{eq:H2}) to the case of a general perfect fluid:
\be
\frac{H^2}{\mbi^2 n^2}=\frac{3\Omega_0(M_1^2-1)-\Omega_1(M_0^2-1)}{6\Omega_1\left[1-\frac32(\rhob+\pb)\Big(\partial_{\rhob} \log\Omega_1+c_s^2 \partial_{\pb} \log\Omega_1\Big)\right]^2} \ .
\label{eq:H2cs2}
\ee
After deriving the modified Friedman equation determining the background solutions, we will re-derive the equation for the tensor perturbations in this more general scenario. This is straightforward because Eq.(\ref{htequation}) is still valid and we only need to obtain the new expressions for the propagation speed $c_T^2$, the GWs normalisation $\alpha$ and the corresponding friction term $\partial_t\log \alpha$. It is easy to see that the expressions (\ref{eq:defcta}) are still valid because they are simply based on the algebraic relations between the two metrics, i.e., they are algebraic functions of the deformation matrix. This in turn implies that $c_T^2$ and $\alpha$ will remain the same and the only effect of considering a general equation of state will go into the friction term because, in general, the expression (\ref{eq:alpha}) will become
\be
\partial_t\log\alpha=-3H(\rhob+\pb)\Big(\partial_{\rhob}\log\alpha+c_s^2\partial_{\pb}\log\alpha\Big).
\ee
However, for the specific case of the EiBI theory, we find from the general expression (\ref{eq:defcta}) that $\alpha=\vert1+\rhob\vert$ for any equation of state. Since this expression does not depend on $\pb$, we see that the potential dependence on $c_s^2$ of the friction term drops and, thus, the friction term also remains oblivious to the presence of $c_s^2$. We then conclude that the equation for the GWs obtained above will still be valid for the more general scenario considered here.

After our discussion on the modifications introduced by considering a general equation of state, we are ready to analyse the potential role of $c_s^2$ for the stabilisation of the GWs in the non-singular solutions of EiBI gravity. The first important feature to notice is that, since  $c_s^2$ does not enter directly into the equation for the tensor modes but only indirectly by possibly modifying the background cosmology, the bouncing and asymptotically Minkowski solutions occurring at the same values of $\rho$ as those discussed in the precedent sections will still be plagued by instabilities. These solutions are obtained as the finite density values that give a vanishing expansion rate in (\ref{eq:H2cs2}), which can occur for values that either cancel the numerator or give a divergent denominator. In order to clarify our analysis, let us explicitly write the modified Friedman equation in terms of the matter variables as
\be
\frac{H^2}{\mbi^2 n^2}=\frac{F_1}{F_2^2} \ ,
\ee
where we have introduced the definitions
\bea
F_1\equiv&&\frac{8}{3}(1+\rhob)(1-\pb)^2\label{def:F1}\\
&&\times\Big[-2+\rhob+3\pb+2(1-\pb)\sqrt{(1-\pb)(1+\rhob)} \Big]\, ,\nonumber\\
F_2\equiv&&(1-\pb)(4+\rhob-3\pb)+3c_s^2(1+\rhob)(\rhob+\pb)\,\label{def:F2} .
\eea
 We then see that the non-singular solutions with $\rhob=-1$ and $\pb=1$ remain in the general case and will be unaffected by $c_s^2$ (as long as $c_s^2$ is regular). As explained above, the GWs on top of these solutions cannot be regularised by the presence of $c_s^2$ so that the only possibility left is that $c_s^2$ allows for new solutions where the tensor modes are stable. By a simple inspection of the above expression we immediately realise that such a possibility can only take place if $c_s^2$ diverges so that we have $H=0$ for $c_s^2\rightarrow \infty$. This is a worrisome feature since having a divergent $c_s^2$ might render the configuration as unphysical if $c_s^2$ is to be regarded as the sound speed of the fluid, i.e., the propagation speed of the density waves. This is indeed the case for perfect fluids for which the pressure is a function of the density alone, i.e., $p=p(\rho)$. In that case we have that $\dot{p}/\dot{\rho}=\diff p/\diff\rho$ so that $c_s^2$ is indeed the adiabatic sound speed of the inhomogeneous perturbations in the fluid. This simple observation indicates that the new non-singular solutions must be supported by some more contrived fluids. Let us notice that this statement refers to the perturbations in the fluid, i.e., even if the background evolution of the fluid behaves as a perfect or barotropic fluid, the non-singular solutions will require non-adiabatic perturbations with $\delta p\neq c_s^2\delta\rho$.

Since we require a divergent $c_s^2$ near the non-singular solution, we can conveniently parameterise it as
\be
c_s^2=c_{s,0}^2+n\mbi\frac{c_{s,1}^2}{H}.
\ee
Near the bouncing and asymptotically Minkowski solutions characterised by $H=0$ at a finite density, we see that $c_s^2$ diverges and, thus, this is a good parametrisation to study the relevant effects around the solutions of interest. Let us notice that $c_{s,0}^2$ and $c_{s,1}^2$ can be functions of time with the only condition that they must remain finite at $H=0$, i.e., the divergent part of $c_s^2$ has been explicitly extracted. In particular, this parametrisation captures the behaviour of a general Lagrangian giving rise to an energy momentum tensor of a perfect fluid form (as a minimally coupled scalar field). In that case, we define the usual time-dependent equation of state parameter $w\equiv p/\rho$ and then we have that
\be
c_s^2\equiv\frac{\dot{p}}{\dot{\rho}}=w\left(1-\frac{\dot{w}}{3Hw(1+w)}\right) \ ,
\ee
which is of the form of our parametrisation, assuming that both $w$ and its time-derivative remain regular. Let us also notice that this simple example shows that the divergence of $c_s^2$ is triggered by the vanishing of $H$. An exception is precisely the case with constant equation of state parameter for which we have the usual relation $c_s^2=w$.

The modified Friedman equation (\ref{eq:H2cs2}) for this case can be written as
\be
\frac{H^2}{\mbi^2 n^2}=\left[\frac{\pm \sqrt{F_1}+3(1+\rhob)(\rhob+\pb)c_{s,1}^2}{F_{2,0}^2}\right]^2 \ ,
\label{eq:H2cs22}
\ee
with $F_1$ given in (\ref{def:F1}) and $F_{2,0}$ the same function as in (\ref{def:F2}) but replacing $c_s^2$ by the regular part $c_{s,0}^2$. We see how the presence of the $1/H$ term in $c_s^2$ generates a new solution of the equation $H=0$ determined by
\be
\pm \sqrt{F_1}+3(1+\rhob)(\rhob+\pb)c_{s,1}^2=0.
\ee
These will be the solutions with some hope of regularising the tensor perturbations. For $c_{s,1}^2=0$ we recover the bouncing and asymptotically Minkowski solutions discussed above. However, despite encountering new potential non-singular cosmological solutions, we need to keep in mind that this comes in at the expense of having a divergent adiabatic sound speed so that, if one of these solutions is found, a crucial viability condition that needs to be carefully checked is the behaviour of the matter sector perturbations around such solutions.

\section{Non-singular solutions and GWs in more general Born-Infeld gravity}\label{moregenBI}

After discussing the non-singular cosmological solutions and the corresponding behaviour of the GWs for the EiBI theory we will turn to analogous analyses within some other specific Born-Infeld inspired theories\footnote{The property of ghost freedom in the EiBI theory is actually generic in metric-affine theories with Lagrangians built out of curvatures where the connection only enters as an auxiliary field. In particular, the theories considered in this section share this property. See  e.g. section 2.7.1 of \cite{BeltranJimenez:2017doy} for more details on this point and \cite{Afonso:2017bxr} for a discussion including non-minimal couplings.}. We will see that many of the properties are actually shared by these theories. In particular the existence of both branches of non-singular cosmologies, namely bouncing and asymptotically Minkowski solutions, also exist for those theories, as expected from the square root structure. We will see how the problems with GWs at the bounce persist for all the cases, while their behaviour can be improved for the asymptotically Minkowski solution of the theory considered in \ref{Sec:polynomial}.

\subsection{Power-law functional extensions}

In this section we shall consider a family of extensions of the original Born-Infeld gravity under the functional form $f(\chi)$, where here we are defining the object $\chi= \det(\Id+\mbi^{-2}\gm^{-1}\mRm(\Gamma))$ by convenience. This way, EiBI gravity corresponds simply to $f(\chi)=\chi^{1/2}$. This class of theories and their cosmological solutions at early times have been extensively discussed in \cite{oor14} (black hole solutions in these theories have been considered instead in Ref.\cite{brw16}). The trick to formulate the corresponding field equations on this case lies on introducing an auxiliary scalar field $A$ so that with the definitions $\phi \equiv df/dA$ and $V(\phi)=A(\phi)f_A -f(A)$, the field equations can be cast as \cite{oor14}

\begin{equation}\label{eq:Rmn-t}
{R^\mu}_\nu(q)=\frac{1}{2 \mpl^2 \phi^2\chi^{3/2}}\left(\LL_G{\delta^{\mu}}_\nu+{T^{\mu}}_\nu\right) \ ,
\end{equation}
where the gravity Lagrangian density is obtained here as $\LL_G\equiv \mbi^2 \mpl^2 (\phi \chi -V(\phi)-\lambda)$, and the auxiliary metric $q_{\mu\nu}$ (compatible with the independent connection, $\nabla_{\lambda}(\sqrt{-q}q^{\mu\nu})=0$) is now defined as in Eq.(\ref{eq:relqg}) now with

\begin{equation}
\hat{\Omega}=\phi \chi^{1/2} (\Id+\mbi^{-2}\gm^{-1}\mRm(\Gamma))
\end{equation}
It should be noted that both $\phi=\phi({T^\mu}_{\nu})$ and $V(\phi)=V(\phi({T^\mu}_{\nu}))$ (as shall be clear with an explicit example below), which means that the corrections on the right-hand-side of the field equations (\ref{eq:Rmn-t}) are just functions of the matter sources, as is common in Palatini theories of gravity.

For the purpose of this work, let us consider the class of power-law theories defined by

\begin{equation} \label{eq:fOmega}
f(\chi)= \chi^\beta \ ,
\end{equation}
where the case $\beta=1/2$ corresponds to the original EiBI theory. From the definitions above, one finds for this case $\phi=\beta \chi^{\beta-1}$ and $V(\phi)=(\beta-1) \chi^{\beta}$ so that the components of the matrix $\hat \Omega = {\rm diag}(\Omega_0,\Omega_0,\Omega_1,\Omega_1)$ read

\begin{equation}
\Omega_0=2\beta \chi^{\beta} w_0 \hspace{0.1cm};\hspace{0.1cm} \Omega_1=2\beta \chi^{\beta} w_1 \ ,
\end{equation}
from where one immediately obtains the components of the matrix $\hat{M}$ as $M_0=\Omega_0^{1/2}$ and $M_1=\Omega_1^{1/2}$. For a perfect fluid (\ref{eq:fluid}) the objects $\{w_0,w_1\}$ read explicitly

\begin{eqnarray}
w_0&=&(1+\rhob +(2\beta-1) \chi^{\beta})^{-1} \label{eq:om1} \\
w_1&=&(1-\pb +(2\beta-1) \chi^{\beta})^{-1} \label{eq:om2}  \ ,
\end{eqnarray}
from where we can obtain the equation

\begin{equation} \label{eq:defOm}
16\beta^4 \chi^{4\beta-1}=\frac{1}{w_0w_1^3} \ .
\end{equation}
Note that, due to the dependence of $\{\omega_1,\omega_2\}$ on $\chi$ in Eqs.(\ref{eq:om1}) and (\ref{eq:om2}), the resolution of this last expression does not provide, in general, a closed relation $\chi=\chi(\rhob,\pb)$ (an exception to this statement is the original EiBI gravity, $\beta=1/2$, for which one obtains the results already discussed in Sec. \ref{sec:BIgravity}). For arbitrary $\beta$ one needs instead to solve Eq.(\ref{eq:defOm}) using numerical methods in order to plug the result back into the field equations (\ref{eq:Rmn-t}) and obtain the evolution of the Hubble factor in these theories. Nonetheless, for the computation of the sound speed and the friction term $\dot{\alpha}/\alpha$ we just need to inspect the relation between the two metrics given by Eq.(\ref{eq:relqg}) and the definitions above to identify the factors $\Omega_0=2\beta^2 \chi^{\frac{4\beta-1}{2}}w_1$ and $\Omega_1=2 \beta^2 \chi^{\frac{4\beta-1}{2}}w_2$, so that from Eq.(\ref{eq:defcta}) we obtain

\begin{eqnarray}
c_T^2&=& \left\vert \frac{1-\pb+(2\beta-1) \chi^{\beta}}{1+\rhob+(2\beta-1) \chi^{\beta}} \right\vert \label{cTpl} \\
\alpha&=& 2\beta^2 \chi^{\frac{4\beta-1}{2}} \left[\frac{1+\rhob+(2\beta-1) \chi^\beta}{(1-\pb+(2\beta-1) \chi^\beta)^3} \right]^{1/2} \label{dapl} \ .
\end{eqnarray}
and, from the last equation, using the conservation law (\ref{eq:continuity}) the explicit form of the friction term $\dot{\alpha}/\alpha$ is immediately obtained. Note that these expressions reduce to those of the original EiBI gravity, Eqs.(\ref{eq:cT}) and (\ref{eq:alpha}), when $\beta=1/2$.

\begin{figure}[t]
\includegraphics[width=8.5cm]{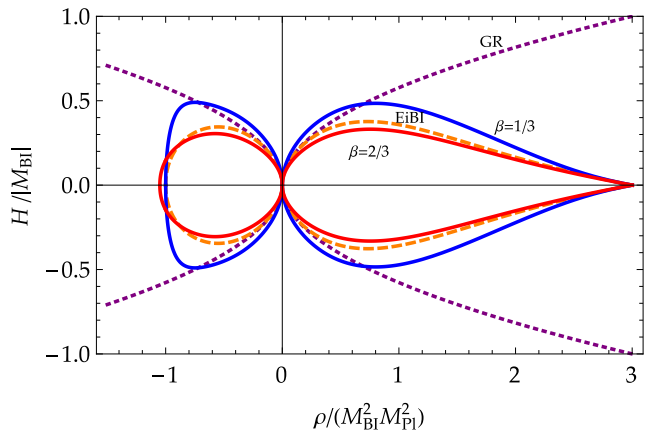}~~~
\caption{We show the dependence of $H$ as a function of $\rho$ for a radiation fluid ($w=1/3$) within the original EiBI gravity (dashed orange line) and for the functional extensions defined by $\beta=1/3$ (solid blue line) and $\beta=2/3$ (solid red line). We see that at low energy densities both recover the usual GR result (dotted purple line). At high energy densities we find instead bouncing solutions (for $\mbi^2>0$) taking place at $\rho_b=-\mbi^2\mpl^2$ for $\beta=1/3$ but at a slightly lower density for $\beta=2/3$, namely, $\rho_b \simeq -1.049 \mbi^2\mpl^2$,  while the asymptotically Minkowski solutions (for $\mbi^2<0$) take place at $\rho_M=3\mbi^2\mpl^2$ in all cases. \label{eq:Fig2}}
\end{figure}

\begin{figure*}
\includegraphics[height=5.8cm,width=8.3cm]{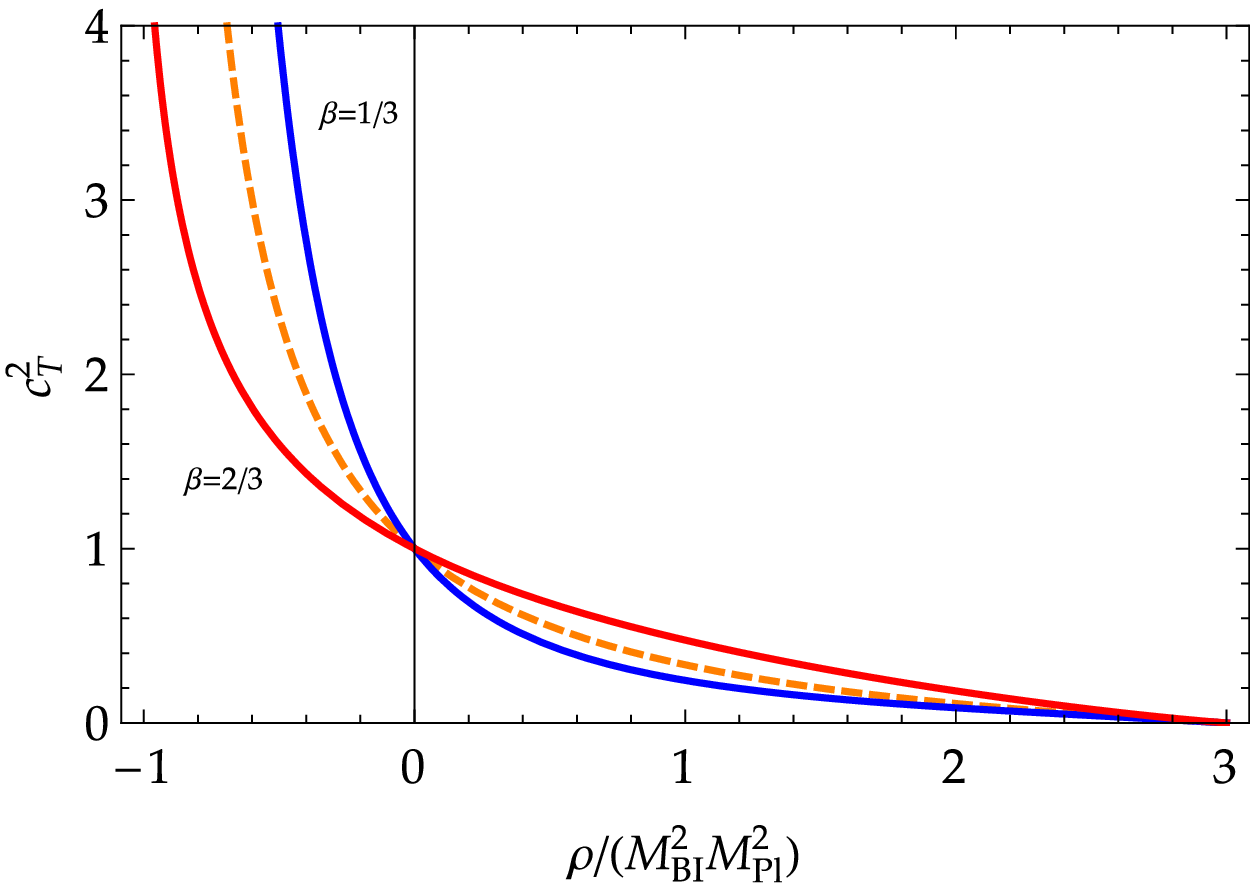}~~~
\includegraphics[height=5.8cm,width=8.3cm]{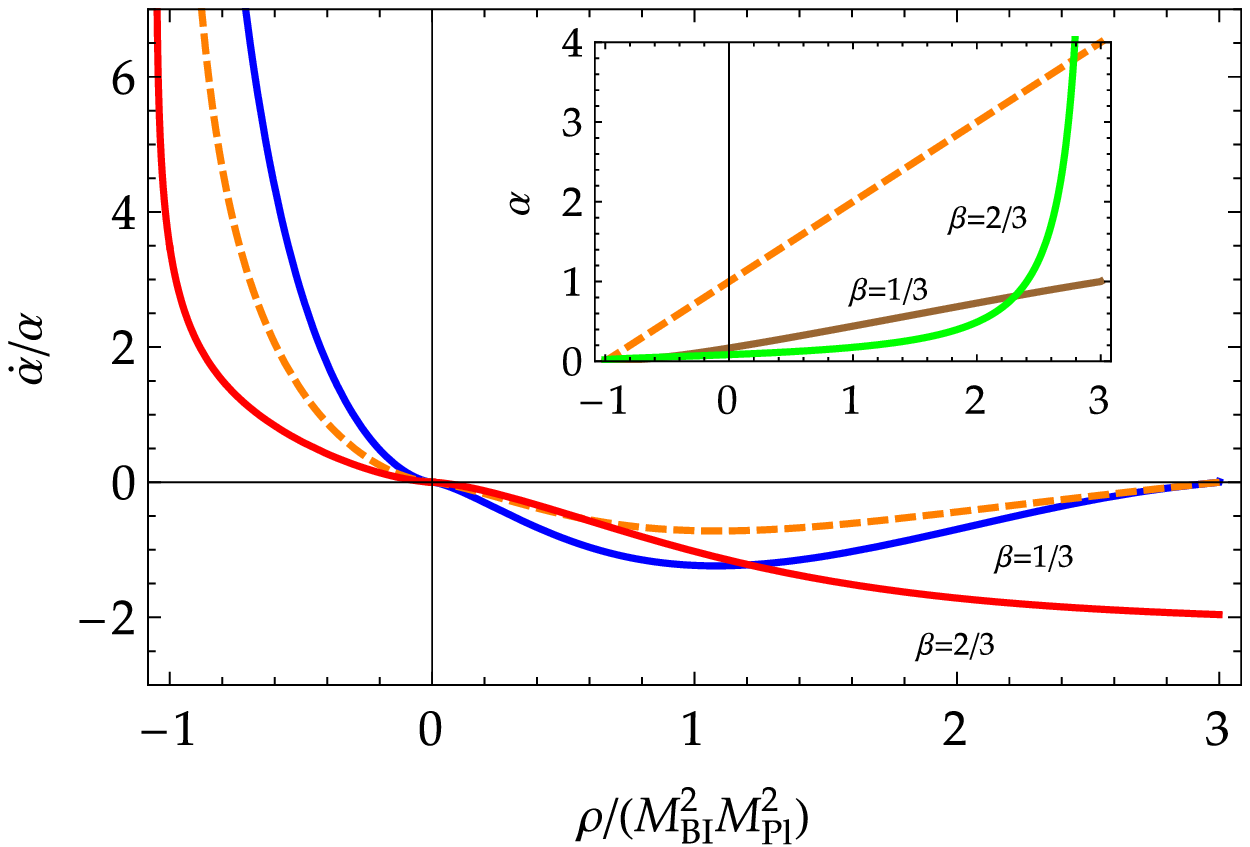}~~~
\caption{{\bf Left panel:} We show the behaviour of the propagation speed, $c_T^2$ (in units of $\vert\mbi\vert$) described by Eq.(\ref{cTpl}) for the functional extensions defined by $\beta=1/3$ (solid blue) and $\beta=2/3$ (solid red).  The dashed orange line shows the behaviour for the original EiBI gravity. For these extensions $c_T^2$ diverges at the bouncing solutions but vanishes at the asymptotically Minkowski ones, like in the EiBI case.
{\bf Right panel:}  We show the behaviour of the friction term, $\dot{\alpha}/{\alpha}$, (in units of $\vert\mbi\vert$) described by Eq.(\ref{dapl}) for the functional extensions defined by $\beta=1/3$ (solid blue) and $\beta=2/3$ (solid red).  The dashed orange line shows the behaviour for the original EiBI gravity. In this case, for the bouncing solutions the friction term diverges (like in EiBI case), but for the asymptotically Minkowski one it becomes finite for the case $\beta=2/3$, i.e., $\dot{\alpha}/{\alpha} \simeq -1.959\vert\mbi\vert$ at $\rho=3\mbi^2\mpl^2$. In the inset plot we also show the normalisation factor of GWs for EiBI and for the two same values of $\beta$ above. \label{eq:Fig3}}
\end{figure*}

In Fig. \ref{eq:Fig2} we have numerically integrated and plotted the evolution of the Hubble factor as a function of the density $\rho$ assuming radiation, $w=1/3$, for the functional extensions defined by $\beta=1/3$ (solid blue) and $\beta=2/3$ (solid red), compared to the original EiBI gravity solution ($\beta=1/2$, dashed orange) and the GR behaviour (dotted purple). There we observe the recovery of the GR solution at low energy densities, while for high energy densities we find instead the same two kinds of solutions as in EiBI gravity. On the one hand we have bouncing solutions ($\mbi^2<0$) which, as it can be explicitly verified from inspection of the different branches of solutions, correspond to $\rho_b=-\mbi^2\mpl^2$ for those extensions with $\beta \leq 1/2$ (for these cases $\chi$ vanishes at the bounce) but to $\rho_b<-\mbi^2\mpl^2$ for those with $\beta>1/2$. In particular, for the $\beta=2/3$ case depicted in this plot, this density reads explicitly $\rho_b \simeq -1.049 \mbi^2\mpl^2$. On the other hand, we find the asymptotically Minkowski solutions ($\mbi^2>0$), which in these extensions occur at $\rho_M=3\mbi^2\mpl^2$ regardless the value of $\beta$.

In Fig. \ref{eq:Fig3} we also employ the functional extensions $\beta=1/3$ (solid blue) and $\beta=2/3$ (solid red). In these cases, at the bouncing solution (corresponding to $\rho_b=-\mbi^2\mpl^2$ for those extensions defined by $\beta \leq 1/2$ but to $\rho_b<-\mbi^2\mpl^2$ for those with $\beta>1/2$), both the propagation speed $c_T^2$ (left panel) and the friction term $\dot{\alpha}/{\alpha}$ (right panel) diverge, which is the same behaviour as in the standard EiBI gravity (dashed orange). On the other hand, for the asymptotically Minkowski solution, $\rho_M=3\mbi^2\mpl^2$, the propagation speed $c_T^2$ vanishes in all cases, while the friction term $\dot{\alpha}/{\alpha}$ vanishes also for those cases defined by $\beta \leq 1/2$, but is finite for those corresponding to $\beta>1/2$ (for the $\beta=2/3$ depicted in this plot it takes the value $\dot{\alpha}/{\alpha} \simeq -1.959\vert\mbi\vert$. These results imply that, like in the standard EiBI gravity ($\beta=1/2$, dashed orange curve), tensorial instabilities have much less dramatic effects in the asymptotically Minkowski solution than in the bouncing one.

\subsection{Elementary symmetric polynomials}\label{Sec:polynomial}

\begin{figure*}
\includegraphics[width=9cm]{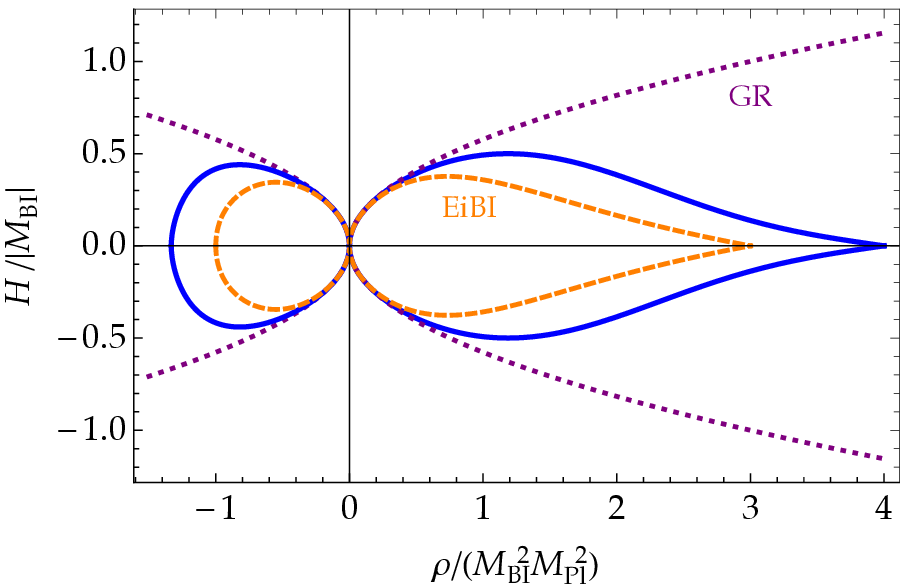}~~~
\includegraphics[width=8.2cm]{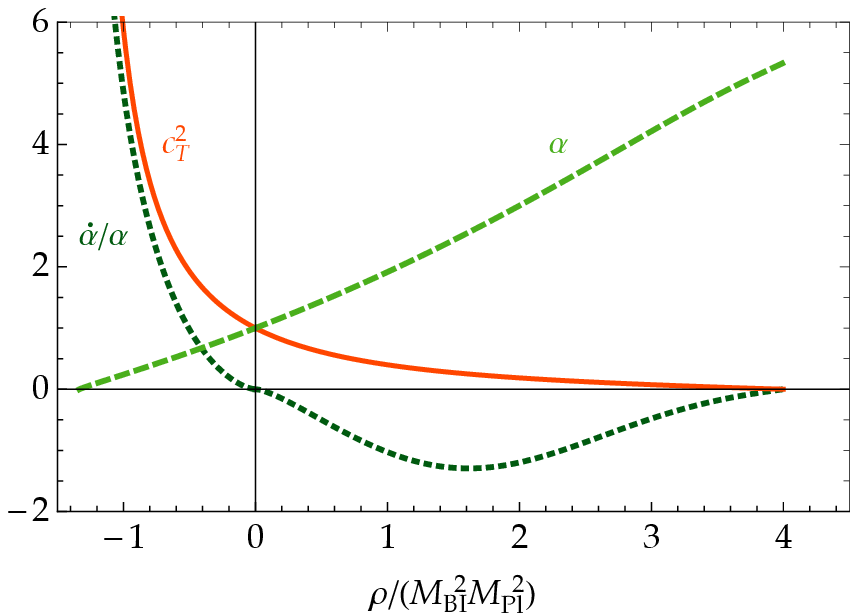}~~~
\caption{{\bf Left panel:} We show the modified Friedman equation (blue) for the theory described by the action (\ref{Eq:beta3action}) and in the presence of a radiation fluid ($w=1/3$). We can see the same qualitatively behaviour as in the EiBI theory (dashed orange line) and, thus, analogous non-singular solutions at high densities, i.e., a bounce for $\mbi^2<0$ and an asymptotically Minkowski solution for $\mbi^2>0$. In both solutions, the maximum allowed density is slightly smaller for the EiBI theory. At low energy densities we recover the usual GR result (dotted purple line). {\bf Right panel:} We show the behaviour of the propagation speed (red solid), the normalisation factor (light-green dashed) and the friction term normalised to $\vert\mbi\vert$ (dark-green dotted)  for the evolution of the GWs in the theory described by (\ref{Eq:beta3action}). We can see an analogous behaviour to the EiBI theory near the bouncing solution, occurring  in this case at $\rho=-4/3 \mbi^2\mpl^2$ and the asymptotically Minkowski solution at $\rho= 4\mbi^2\mpl^2$.\label{Fig:H2beta3}}
\end{figure*}

Let us now consider an extended version of the Born-Infeld theory with similar solutions as those above, but in which the instability of the tensor modes for the asymptotically Minkowski solution is avoided, while the divergence for the bouncing solution remains. Such an extension was considered in \cite{Jimenez:2014fla} inspired by the massive gravity potential interactions and consisted in extending the EiBI action to all the elementary symmetric polynomials of the matrix $\Mm$. The cosmology of the first polynomial was already studied in \cite{Jimenez:2014fla} where a de Sitter phase at high densities was identified and used in \cite{Jimenez:2015jqa} to develop an inflationary phase supported by a dust component. Here we will focus on the third polynomial which, as we will show, has non-singular solutions analogous to the ones discussed above for the EiBI model\footnote{The first two polynomials do not exhibit these solutions for the simplest matter sources and the fourth polynomial is nothing but the EiBI model, which is why we focus on the third polynomial here. On the other hand, the symmetric properties of the elementary polynomials lead to simpler equations than in more general theories like e.g. the ones considered in the previous section.}. Thus, the action that we will consider is given by
\be
\mS_3=\frac{\mbi^2\mpl^2}{18}\int\diff^4x\sqrt{-g}\Big( [\Mm]^3- 3[\Mm][\Mm^2]+2[\Mm^3] -\lambda\Big)
\label{Eq:beta3action}
\ee
where $[\cdot]$ stands for the trace of the corresponding matrix and the normalisation has been chosen so that we recover GR at low curvatures. We will additionally generate a cosmological constant which can be cancelled by choosing $\lambda=4/3$. We will assume this value in the following and consider the cosmological constant as part of the matter sector. The metric field equations in this case are given by
\bea
\frac{3M_1^2}{M_0}+M_1^3&=&4+3\rhob,\\
2M_0+M_1+M_0 M_1^2&=&4-3\pb.
\label{eq:M0M1EBI}
\eea
The solutions of these equations are more complicated than in the EiBI case. In particular, we find four different branches of solutions. A detailed treatment of such solutions is beyond the scope of this paper and we will restrict to the branch that is continuously connected with GR at low energy-densities. The existence of this branch is guaranteed by the fact that our action reduces to Einstein-Hilbert in the regime of small curvatures.

The Hubble expansion rate takes the same form as in (\ref{eq:H2}), but now $M_0$ and $M_1$ are the solutions of (\ref{eq:M0M1EBI}) and the components of the deformation matrix relate to those of the fundamental matrix $\hat{M}$ by
\bea
&&\Omega_0=\sqrt{\frac{M_0}{27M_1^2}\big(2M_0+M_1\big)^3}\,,\\
&&\Omega_1=\sqrt{\frac{M_1^2}{3M_0}\big(2M_0+M_1\big)}\,.
\eea
The dependence of the deformation matrix on the density and pressure of the matter source is determined by the metric field equations given in (\ref{eq:M0M1EBI}). The explicit expression is not very illuminating in this case so we will not reproduce it here. However, the modified Friedmann equation giving the Hubble expansion rate as a function of the density and pressure can be easily plotted. Since the aim of the present work is not to perform an exhaustive analysis of this type of cosmological solutions for this theory, but rather explore the behaviour of the GWs around non-singular solutions we will focus on the case of a radiation dominated universe and, as explained above, we only consider the branch of solutions continuously connected with GR. For that case, we plot the Hubble expansion rate as a function of the density in Fig. \ref{Fig:H2beta3}, where we can see that this class of Born-Infeld gravity exhibits the same non-singular solutions as the EiBI model, i.e., we have an asymptotically Minkowski solution and a genuine bouncing solution. This shows once again that these solutions are typical of Born-Infeld inspired gravity theories. The two values of $\rho$ at which the Hubble expansion rate vanishes are $\rho_b=-4/3\mbi^2\mpl^2$ and $\rho_M= 4 \mbi^2\mpl^2$ for the bouncing and the asymptotically Minkowski solutions respectively.

The evolution of the GWs near these solutions is again determined by the behaviour of the propagation speed and the friction terms, which in this case are given by
\bea
&&c_T^2=\left\vert\frac{M_0}{3M_1^2}\big(2M_0+M_1 \big),   \right\vert\\
&&\alpha=\frac{M_1^2}{M_0}.
\eea
The dependence of both the propagation speed and the friction term is shown in Fig. \ref{Fig:H2beta3}. In that figure we can clearly see that the pathological behaviour of the GWs in the bouncing solutions is also present in this theory, i.e., both $c_T^2$ and $\dot{\alpha}/\alpha$ diverge. The situation is also analogous for the asymptotically Minkowski solution where the propagation speed and the friction term vanish, but the normalisation remains strictly positive and finite, indicating that the dispersion relation is expected to be determined by operators of higher order in derivatives.

\section{Conclusions}\label{conclusions}

In this work we have revisited the issue of tensor instabilities for non-singular cosmological solutions within the framework of Born-Infeld inspired theories of gravity. We have briefly reviewed the existence of bouncing and asymptotically Minkowski solutions in the EiBI model and the appearance of instabilities in the GWs sector for such solutions. We have re-obtained known results in a slightly different approach that allows an apparent understanding of these instabilities without even solving the background evolution. We have discussed how GWs are sensitive to a different geometry from that to which matter fields couple minimally and it is precisely the singular behaviour of the metric seen by GWs what causes the pathologies, even if the matter fields see a perfectly regular geometry. We have reduced the properties of the GWs to the behaviour of the anomalous propagation speed, the normalisation and the friction term generated by the deformation matrix relating the two geometries. The fact that the bouncing solutions give rise to divergences in the propagation speed and a vanishing normalisation explains the pathology of the GWs evolution at the bounce as a strong coupling problem. On the other hand, the asymptotically Minkowski solution leads to a vanishing value for the propagation speed, what implies a much milder type of instability which, in fact, could be resolved by including higher derivative terms, in analogy with the ghost condensate model. This has been briefly discussed at the homogeneous but non-linear level, where we have seen that the bouncing solutions result in a divergence of the shear at the bounce, while it behaves smoothly in the asymptotically Minkowski solution.

A remarkable property of modified gravity theories within the metric-affine formalism is the sensitivity of the background cosmology to the sound speed of the matter source. In GR, only the equation of state parameter is relevant so this represents a very distinctive feature. We have thus studied the potential impact of this new parameter on the non-singular solutions and the possibility of regularising the GWs behaviour. We have found that such a parameter does not affect the anomalous propagation speed of GWs and, therefore, the presence of a non-trivial sound speed for the matter sector supporting the non-singular solutions cannot regularise the pathological behaviour of GWs. Another possibility is that the sound speed can generate new non-singular solutions and these could be free from instabilities. Unfortunately, we found that such solutions must rely on a divergent sound speed, so the physical viability of such solutions concerning the perturbations of the matter sector is jeopardised. This shows that new non-singular solutions with no  instabilities, both in the GWs and the matter sector, if they exist, they must be supported by contrived matter sources. In particular, sources that can be described by perfect fluids can hardly achieve it.

After studying the tensor instabilities within the EiBI theory, including the non-trivial effects of the matter sector sound speed, we have considered some extensions of this theory, in particular power law and elementary symmetric polynomials extensions, where the same non-singular solutions can be found. We have seen that the power law extensions exhibit similar properties to the EiBI theory, i.e., the bouncing solutions come in at the expense of divergent propagation speed and friction terms for the GWs equation, while the asymptotically Minkowski solutions lead to a vanishing propagation speed. The vanishing or not of the friction term depends on a new parameter defining a particular theory. We have finally shown the same behaviour appearing in the elementary symmetric polynomials extension.
%

In order to summarise, we could say that the bouncing solutions for the class of Born-Infeld inspired theories studied throughout this work are prone to the presence of severe instabilities in the gravitational waves sector. Our findings are in no way a definite no-go result for the existence of fully stable bouncing solutions within these theories, but they show that the simplest realisations are not possible and, in general, finding stable bouncing solutions is challenging, as in many other frameworks. On the other hand, the instabilities found in the branch of asymptotically Minkowski solutions seem less severe than the ones of the bouncing solutions because the normalisation remains finite and non-vanishing. Our results might signal that the two classes of non-singular solutions existing within Born-Infeld type of theories come in at the expense of a vanishing propagation speed and, furthermore, they might admit a characterisation provided by the behaviour of the normalisation factor so that bouncing and asymptotically Minkowski solutions correspond to vanishing and finite normalisation factors respectively. The generality of these results remains to be established.

\section*{Acknowledgments}
We are grateful to Daniel de Andr\'es Hern\'andez for pointing out a mistake in a previous version that changed the conclusions in Sec \ref{Sec:polynomial}.
J.B.J. acknowledges the financial support of A*MIDEX project (No.~ANR-11-IDEX-0001-02) funded by the Investissements d'Avenir French Government program, managed by the French National Research Agency (ANR), MINECO (Spain) projects FIS2014-52837-P, FIS2016-78859-P (AEI/FEDER) and Consolider-Ingenio MULTIDARK CSD2009-00064. L.H. is funded by Dr. Max R\"ossler, the Walter Haefner Foundation and the ETH Zurich Foundation. G. J. O. is supported by a Ramon y Cajal contract and the grant FIS2014-57387-C3-1-P from MINECO and the European Regional Development Fund. D. R.-G. is funded by the Funda\c{c}\~ao para a Ci\^encia e a Tecnologia (FCT, Portugal) postdoctoral fellowship No.~SFRH/BPD/102958/2014 and the FCT research grant UID/FIS/04434/2013. This work is also supported by the Consolider Program CPANPHY-1205388, the Severo Ochoa grant SEV-2014-0398 (Spain), and the CNPq (Brazilian agency) project No.~301137/2014-5. This article is based upon work from COST Action CA15117, supported by COST (European Cooperation in Science and Technology).


\end{document}